\documentclass[twocolumn,aps,prb,longbibliography,superscriptaddress,floatfix]{revtex4-2}
\usepackage{nicefrac}
\usepackage{amsfonts}
\usepackage{mathtools,amssymb,graphicx,units,bm}
\usepackage{amsmath}
\usepackage{subfigure}
\usepackage{multirow} 
\usepackage{tabularx} 
\usepackage{array}

\usepackage[plainpages=false,pdfpagelabels,colorlinks=true,linkcolor=red,urlcolor=PineGreen,citecolor=PineGreen,pdftitle={Title},pdfauthor={},pdfdisplaydoctitle=true,pdfduplex=DuplexFlipLongEdge]{hyperref}

\usepackage{soul} 
\usepackage[dvipsnames,usenames]{color}
\usepackage[normalem]{ulem}
\graphicspath{{figures/}}
\usepackage{color}
\usepackage{bbold} 
\usepackage{float}
\usepackage{silence}
\usepackage{dcolumn}
\usepackage{physics}%
\usepackage[dvipsnames]{xcolor}
\emergencystretch=\maxdimen
\DeclareMathOperator{\sign}{sign}

\usepackage{units}
\usepackage{tensor} 
\usepackage{braket}
\usepackage{hyperref}
\usepackage{amsfonts}
\usepackage{fontenc}
\usepackage{graphicx}
\usepackage{xcolor}
\usepackage{textcomp}
\usepackage{epstopdf}
\usepackage{braket}
\usepackage{mathtools}
\usepackage{amsmath}
\usepackage{dcolumn}
\usepackage{multirow}
\usepackage{units}
\usepackage{ulem}
\usepackage{mathrsfs, amssymb, bm}

\begin{document} 

\title{Topological Phases in Fractals: Local Spin Chern Marker in the Sierpinski carpet Kane-Mele-Rashba Model}

\author{L. L. Lage}
\email{lucaslage@id.uff.br}
\affiliation{Instituto de F\'isica, Universidade Federal Fluminense, Niter\'oi, Av. Litor\^{a}nea sn 24210-340, RJ-Brazil}

\author{A. B. Félix}
\affiliation{Instituto de F\'isica, Universidade Federal Fluminense, Niter\'oi, Av. Litor\^{a}nea sn 24210-340, RJ-Brazil}

\author{S. dos A. Sousa-Júnior}
\affiliation{Department of Physics, University of Houston, Houston, Texas 77204, USA}

\author{A. Latg\'e}
\affiliation{Instituto de F\'isica, Universidade Federal Fluminense, Niter\'oi, Av. Litor\^{a}nea sn 24210-340, RJ-Brazil}

\author{ Tarik P. Cysne}
\email{tarik.cysne@gmail.com}
\affiliation{Instituto de F\'isica, Universidade Federal Fluminense, Niter\'oi, Av. Litor\^{a}nea sn 24210-340, RJ-Brazil}

\date{\today}

\begin{abstract}
We study the spectral properties and local topology of the Kane-Mele-Rashba model on a Sierpinski Carpet (SC) fractal, constructed from a rectangular flake with an underlying honeycomb arrangement and open boundary conditions. When the system parameters correspond to a topologically trivial phase, the energy spectrum is characterized solely by bulk states that are not significantly modified by the system's fractality. For parameters corresponding to the quantum spin Hall insulator (QSHI) phase, in addition to bulk states, the energy spectrum exhibits in-gap topological states that are strongly influenced by the fractal geometry. As the fractal generation increases, the energy spectrum of the in-gap topological states develops a staircase-like profile, resulting in sharp peaks in the density of states. We also show that both the QSHI and the trivial phase exhibit a large gap in the valence-projected spin spectrum, allowing the use of the local spin Chern marker (LSCM) to index the local topology of the system. Fractality does not affect this gap, allowing the application of LSCM to higher fractal generations. Our results explore the LSCM versatility, showing its potential to access local topology in complex geometries such as fractal systems.
\end{abstract}

\maketitle


\section{Introduction}\label{Intro}

\begin{figure*}[t]
    \centering
    \includegraphics[width=0.98\textwidth]{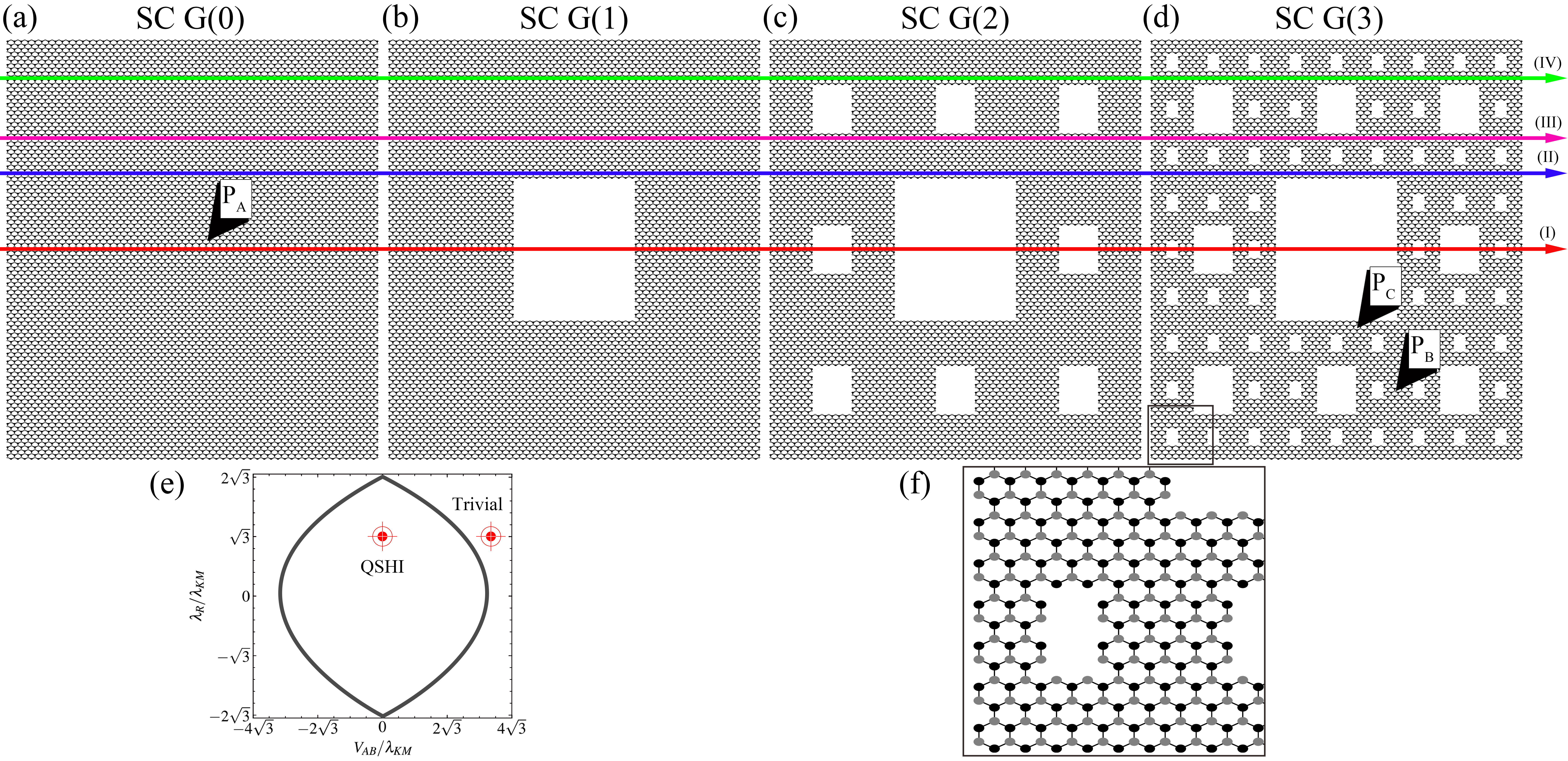}
   \caption{\textcolor{black}{(a-d)} Sierpinski Carpets (SCs) with a honeycomb mesh for the first three fractal orders. Colored arrows (I)–(IV) indicate paths for LSCM calculations in Fig. \ref{LSCM-lines-QSHI-trivial}. The number of sites in each generation is $N=11988$ [G(0)], $N=10512$ [G(1)], $N=9456$ [G(2)], and $N=8688$ [G(3)]. \textcolor{black}{(e)} Schematic of the topological phase diagram for the Kane-Mele-Rashba model, with red targets marking parameter sets used in Figs. \ref{figenergydos}-\ref{site-resolved} for QSHI and trivial phases. \textcolor{black}{(f)} Magnified view of the honeycomb mesh used in the fractal construction. \textcolor{black}{In panel (a), P$_{\rm A}$ is the lattice site used to calculate the phase diagram presented in Fig. \ref{figDiagram}(a). In panel (d), P$_{\rm B}$ and P$_{\rm C}$ are the lattice site used to calculate the phase diagram presented in Fig. \ref{figDiagram}(b, c).}}
   \label{fig1}
\end{figure*}

Fractal geometries, characterized by self-similarity and non-integer dimensionality, have emerged as a rich playground for exploring the interplay between geometry, topology, and quantum mechanics. Among these, the structure of a Sierpinski fractal, recursively constructed with a hierarchical geometry, has garnered significant attention due to its unique spectral properties and potential applications in condensed matter physics with recent experimental realizations \cite{Shang2015,Kempkes2019,LI20222040} and theoretical investigations \cite{Yang2020,D2CP02426H,Lage_2025,Osseweijer2024, Iliasov2020, Li2023, vanVeen2017, Yang2020, Manna2022, Manna2023, Manna2020}. In particular, the Sierpinski carpet (SC) is a self-similar fractal system with a Hausdorff dimension of $ \log_3(8) \approx 1.8928 $. It is constructed through an iterative process of subdividing a solid square into a $ 3 \times 3 $ grid of smaller congruent squares and removing the central square. This process is repeated recursively for each of the remaining eight squares, resulting in a structure composed of self-replicated solid squares, as shown in Fig. \Ref{fig1}. Studies have explored the consequences of fractality in charge and spin transport and the Hofstadter spectrum of electrons in SC fractal geometry \cite{Brzezinska2018, Iliasov2020, vanVeen2016, Sandoval2025, Fischer2021}.

In recent years, the field of topological phases of matter has witnessed remarkable progress, driven by the discovery of topological insulators and superconductors. In the band-theory perspective [systems in the thermodynamic limit with translation symmetry] the topological phases are characterized by bulk topological invariants that index the global properties of a system \cite{Hasan2010}. These invariants are protected by a gapped bulk energy spectrum and remain unchanged under smooth variations in the electronic Hamiltonian. Non-trivial topology presents physically measurable consequences. For example, the Chern insulators proposed by Haldane \cite{Haldane1988} are indexed by a topological invariant called the Chern number and are characterized by a quantized Hall conductivity. This conductivity is enabled by the emergence of edge states when the system is prepared, for instance,  in a nanoribbon geometry. Similarly, the quantum spin Hall insulators (QSHI) introduced by Kane and Mele \cite{Kane2005a, Kane2005b} are characterized by a finite spin-Hall conductivity within the insulating bulk energy gap. This conductivity is enabled by spin-polarized edge states that appear in a nanoribbon geometry. QSHIs can be indexed by a $\mathbb{Z}_2$ invariant protected by time-reversal symmetry. However, in systems with broken translational invariance, such as fractals, quasicrystals, or disordered lattices the global topological indexation may lose its meaning, requiring a local description of topological properties. Added to this, from a fundamental point of view, studying topological aspects of fractals \cite{Salib2024, Eek2024, Osseweijer2024} is appealing because their non-integer Hausdorff dimension does not fit into the ten-fold way topological classification \cite{Altland1997, Ryu2010}. 

Local markers have emerged as a powerful tool for addressing the challenge of topological indexing systems that lack translational symmetry. By providing a real-space measure of topological invariants, local markers enable the characterization of topological phases in systems with complex geometries or a high degree of disorder \cite{Assuncao2024, Oliveira2024, Kim2023, Drigo2020}. It has also been used to study strongly correlated topological systems in different models \cite{Melo2023, Gilardoni2022, Markov2021}. This approach is particularly relevant for Sierpinski fractals in which the interplay between fractal geometry and topology can lead to novel phenomena, such as hierarchical edge states \cite{Lage_2025}, self-similar energy spectra \cite{Pedersen2020, Domany1983, Rammal1982}, and fractality-induced topological phases \cite{Eek2024}. 

Experimental setups in bismuth fractal flakes were recently investigated and highlighted the importance of including Rashba spin-orbit coupling (RSOC) effects in describing such systems \cite{Canyellas2024}. The RSOC is associated with the breaking of $(z\rightarrow -z)$-symmetry produced by the proximity of materials to a substrate and causes the electronic Hamiltonian to break spin conservation,  $[\mathcal{H},\hat{s}_z]\neq 0$. A local marker description of the QSHI phase for systems with strong RSOC was recently developed \cite{Bau2024}. The local spin Chern marker (LSCM) combines the local Chern number introduced by Bianco and Resta \cite{Bianco2011} with Prodan’s spin Chern number approach, which was constructed in Ref. \cite{Prodan2009} assuming the thermodynamic limit. The LSCM was applied to describe the local topology of generalized classes of Kane-Mele-Rashba models \cite{tiao}, demonstrating its versatility as a tool for real-space topological indexation.

Motivated by the previous discussion, we investigate the local topology of the Kane-Mele-Rashba model in the SC geometry using the LSCM. Studies of SC geometry in an underlying honeycomb mesh have been motivated by potential applications in graphene-based fractal nanostructures \cite{Yang2022, Bouzerar2020, vanVeen2016, Han2020, Yang2024}. In principle,  Kane-Mele-Rashba model could be realized in graphene through the proximity effect to a suitable substrate \cite{Cysne2018, Jin2013, Song2018, Gmitra2016, Han2014, Avsar2020, Zollner2025}. Given the maturity of nanofabrication technology for two-dimensional materials, graphene can play an important role in the experimental understanding of the topological nature of \textcolor{black}{electronic} fractals. Other systems in which the Kane-Mele-Rashba model shares similarities include graphene family materials \cite{Ezawa2012, Ezawa2015, Liu2011} and bismuthene grown in SiC \cite{Li2018, Reis2017}.

\section{Model and Methods}
\subsection{Kane-Mele-Rashba Hamiltonian in Sierpinski Carpet \label{SecKMRSC}}

The Kane-Mele-Rashba Hamiltonian can be cast as,
\begin{eqnarray}
\mathcal{H}_{\rm KMR}&=&t\sum_{s=\uparrow,\downarrow}\sum_{\langle {\bf i}, {\bf j} \rangle}c^{\dagger}_{{\bf i},s}c_{{\bf j},s}+V_{\rm ab}\sum_{s=\uparrow,\downarrow}\sum_{\bf i}\tau_{\bf i}c^{\dagger}_{{\bf i},s}c_{{\bf i},s}\nonumber \\
&&\hspace{0mm}+i\lambda_{\rm KM}\sum_{s,s'}\sum_{\langle \langle {\bf i}, {\bf j}\rangle \rangle} \nu_{{\bf i}{\bf j}} c^{\dagger}_{{\bf i},s} \left(\hat{s}_z\right)_{s,s'}c_{{\bf j},s'} \nonumber \\
&&\hspace{0mm} +i\lambda_{\rm R}\sum_{s,s'}\sum_{\langle {\bf i}, {\bf j}\rangle}c^{\dagger}_{{\bf i},s}\left[\left( \mathbf{\hat{s}}\times\mathbf{d}_{{\bf i}{\bf j}} \right)_z \right]_{s,s'}c_{{\bf j},s'}, \label{HKMR}
\end{eqnarray}
where the first term represents the nearest neighbor hopping of electrons, with $c^{\dagger}_{{\bf i},s}$   and $c_{{\bf j},s}$ being the fermionic creation and annihilation operators for electrons at sites ${\bf i}$ and ${\bf j}$ of the lattice and spin $s=\uparrow,\downarrow$; $\langle {\bf i}, {\bf j} \rangle$ indicate a sum running over nearest neighbor sites. In all results presented in this work, we set the hopping amplitude $t$ as the unit of energy [$t=-1$]. The second term is the sublattice potential. $\tau_{\bf i}=+1$ for sites ${\bf i}$ belonging to sublattice A of honeycomb structure, and $\tau_{\bf i}=-1$ for sites ${\bf i}$ belonging to sublattice B of honeycomb structure. The third term is the Kane-Mele spin-orbit coupling that gives rise to the QSHI phase. $\nu_{{\bf i}{\bf j}}=\text{sign}\left({\bf d}_1\times {\bf d}_2\right)_z=\pm 1$, where ${\bf d}_1$ and ${\bf d}_2$ are unit vectors along the two bonds that the electron crosses when jumping from the site ${\bf j}$ to the next nearest neighbor ${\bf i}$ [indicated in Eq.(\ref{HKMR}) by $\langle \langle {\bf i},{\bf j}\rangle \rangle$]. Finally, the fourth term is the Rashba spin-orbit coupling $\mathcal{H}_{\rm R}$. $\mathbf{d}_{{\bf i}{\bf j}}$ is the unity vector along the direction connecting site ${\bf i}$ to site ${\bf j}$, $\mathbf{\hat{s}}=\hat{s}_x{\bf x}+\hat{s}_y{\bf y}+\hat{s}_z{\bf z}$ is related to physical spin of electrons (up to a factor $\hbar/2$), and $\hat{s}_{x,y,z}$ are Pauli matrices. The Rashba coupling breaks $(z\rightarrow -z)$-symmetry \cite{Bychkov-Rashba_1984} and is responsible for the non-conservation of spin in the full Hamiltonian $\left[\mathcal{H}_{\rm R}, \hat{s}_z\right]\neq 0$.

The model described by Eq.(\ref{HKMR}) is applied to the lattices shown in Fig. \ref{fig1}, illustrating different generations of SC.  A rectangular flake corresponds to a zero-order SC and it is shown in Fig. \ref{fig1} (a) [G(0)]. Figs. \ref{fig1} (b)-(d) show the first [G(1)], second [G(2)], and third [G(3)] SC generations, respectively. Open boundary conditions (OBC) are used in the numerical calculation. 
We call attention to an important aspect of the construction presented in Fig. \ref{fig1}: When SC is built on an underlying square lattice, it has a four-fold rotation symmetry [$C_4$]. However, the used SC is based on a honeycomb lattice, and the entire structure exhibits two-fold rotation symmetry [$C_2$] \textcolor{black}{when $V_{\rm ab}=0$}. In fact, it is clear from Fig. \ref{fig1} that while the upper and lower boundaries terminate in zigzag edges, the left and right boundaries are armchair edges. As shown in Ref. \cite{vanVeen2016}, the SC in the underlying honeycomb lattice exhibits typical fractal features and the expected Hausdorff dimension is $d_{\rm H}\approx 1.89$.

\subsection{Local Spin Chern Marker \label{SecLSCM}}

We employ the LSCM \cite{Bau2024, Chen2023b, Wieder-Bradlyn-NatCommun, tiao} to study the local topology of Hamiltonian from Eq.(\ref{HKMR}) in SC lattices. The LSCM is defined by
\begin{eqnarray}
\mathfrak{C}_s({\bf r})=\frac{\mathfrak{C}_+({\bf r})-\mathfrak{C}_-({\bf r})}{2}, \label{SpinChernMarker} 
\end{eqnarray}
where,
\begin{eqnarray}
\mathfrak{C}_{\sigma}({\bf r})=2\pi \text{Im}\bra{{\bf r}}\mathcal{Q}_{\sigma}\hat{X} \mathcal{P}_{\sigma}\hat{Y} \mathcal{Q}_{\sigma} - \mathcal{P}_{\sigma}\hat{X} \mathcal{Q}_{\sigma}\hat{Y} \mathcal{P}_{\sigma} \ket{{\bf r}},
\label{Csigma}
\end{eqnarray}
with $\hat{X}$ and $\hat{Y}$ being the components of the position operators $\hat{{\bf r}}$. 
To construct $\mathcal{Q}_{\sigma}$ and $\mathcal{P}_{\sigma}$, one first needs to build the valence state projectors $\mathcal{P}=\sum^{N_{\text{occ}}}_{n=1}\ket{u_n}\bra{u_n}$, where $\ket{u_n}$ is the eigenstate of the SC lattice model Hamiltonian $\mathcal{H}_{\rm KMR}$ with state index $n=(1,...,2N)$, $N$ is the number of sites while $N_{\rm occ}$ is the number of occupied states. Here, $N_{\rm occ}$ corresponds to a half filling, unless explicitly stated otherwise. Then, we construct the spin matrix projected onto the valence subspace $M^{s_z}_{\text{v.s.}}=\mathcal{P}\hat{s}_z\mathcal{P}$ and perform its diagonalization $M^{s_z}_{\text{v.s.}}\ket{\phi_{v}}=\xi_v^{s_z}\ket{\phi_{v}}$, where ($v=1,..., N_{\text{occ}}$). For time-reversal symmetric models, the $N_{\text{occ}}$ eigenvalues $\xi_v^{s_z}$ are symmetrically distributed around zero within the interval $[-1,1]$. The branches of this spectrum can be separated by the sign of their eigenvalues [$\sigma=+/-$ for positive/negative $\xi_v^{s_z}$, respectively], 
if there is a finite gap $\Delta^{s_z}$. The normalized eigenvectors of $M^{s_z}_{\rm v.s.}$, $\ket{\phi_v}$, has dimension $N_{\text{occ}}$ and can be written as $\ket{\phi_{v}} =\left( \beta_{1,v}, \beta_{2,v},..., \beta_{N_{\rm occ},v}\right)^{\text{T}} $, T meaning the transpose operation. We then define the states,
\begin{eqnarray}
\ket{\psi_{v}} =\sum^{N_{\text{occ}}}_{\alpha=1} \beta_{v,\alpha} \ket{u_{\alpha}}, \label{Rotsigma}
\end{eqnarray}
in terms of the valence eigenstates $\ket{u_{\alpha}}$ of the SC lattice model Hamiltonian $\mathcal{H}_{\rm KMR}$. The projectors used in Eq. (\ref{Csigma}) are given by
\begin{eqnarray}
\mathcal{P}_{\sigma}=\sum_{\xi^{s_z}_v|\sign [\xi^{s_z}_v]=\sigma} \ket{\psi_{v}}\bra{\psi_{v}}. \label{Psigma}
\end{eqnarray}
Finally, the complementary matrices in Eq. (\ref{Csigma}) are defined as $\mathcal{Q}_{\sigma} = \mathbb{1}- \mathcal{P}_{\sigma}$. 
 
\section{Spectral Properties of KMR model in SC}
\subsection{Energy spectra}
\begin{figure}[h]
    \centering
    \includegraphics[width=8.7cm]{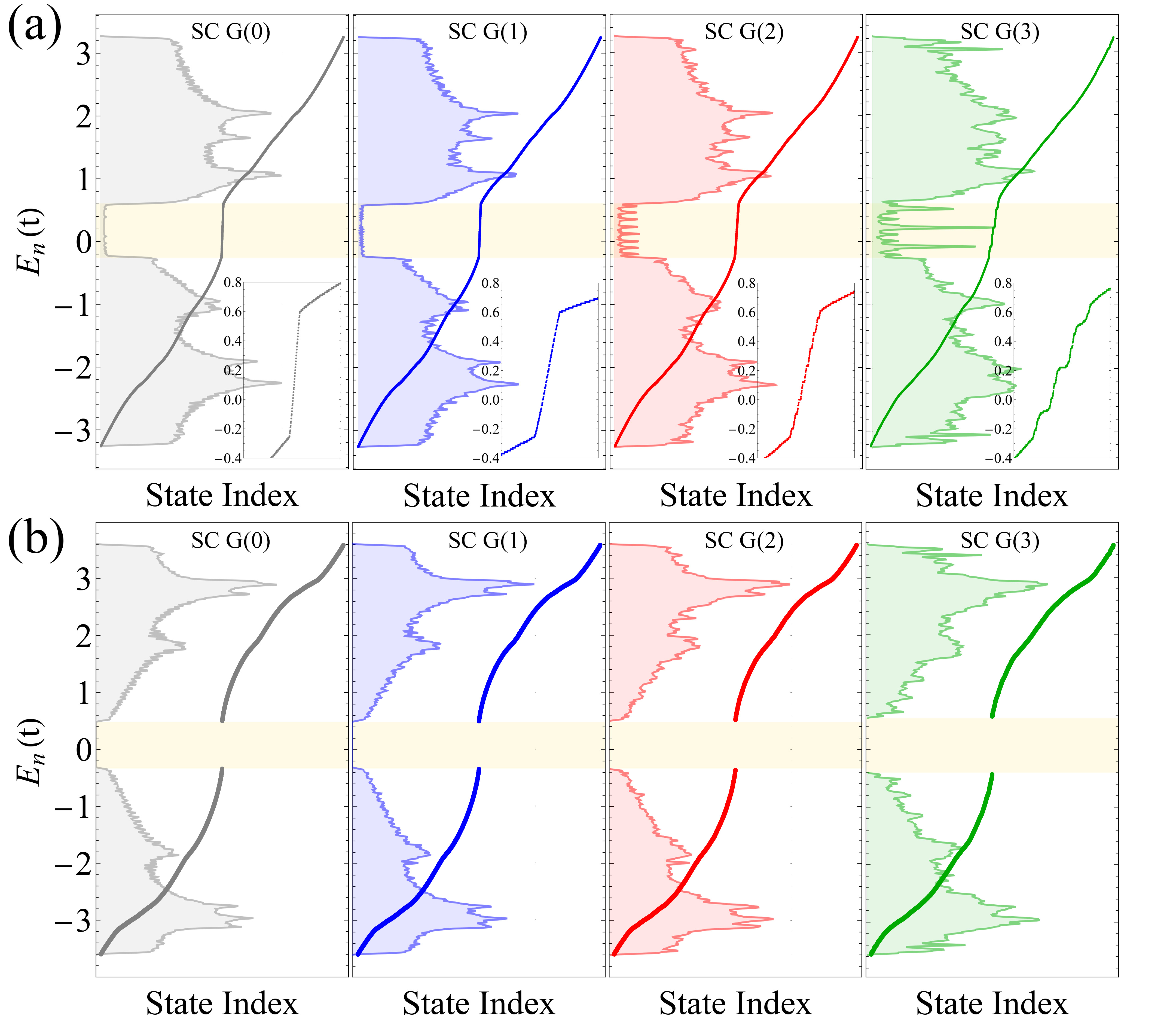}
   \caption{Energy spectra and DOS for $\lambda_{\rm KM}=0.25t$ and $\lambda_{\rm R}=\sqrt{3}\lambda_{\text{KM}}\approx0.43t$ at (a) QSHI ($V_{\text{ab}}=0$) and (b) trivial phase ($V_{\text{ab}}=7\sqrt{3}t/8\approx 1.51t$). Results for G(0-3) appear in gray, blue, red, and green. The light-yellow region marks the bulk band gap ($\Delta^{\rm E}$) of the homogeneous Kane-Mele-Rashba model. Insets in (a) zoom into this region, highlighting the topological states inside the gap with the staircase profile.}
   \label{figenergydos}
\end{figure} 
In Fig. \ref{figenergydos}, we show the energy spectra and density of states (DOS) of the model Hamiltonian $\mathcal{H}_{\rm KMR}$, for the four generations of the SC presented in Fig. \ref{fig1} (a)-(d). Unless explicitly stated otherwise, we consider the model's parameters $\lambda_{\rm KM}=0.25t$, $\lambda_{\rm R}=\sqrt{3}\lambda_{\text{KM}}\approx0.43t$ with $V_{\rm ab}=0$ for the QSHI phase and $V_{\rm ab}=7\sqrt{3}t/8$ for the topologically trivial phase. These sets of parameters are chosen conveniently and marked as the two red targets in the schematic diagram at Fig. \ref{fig1} (e).

\begin{figure}[!h]
    \centering
    \includegraphics[width=8.cm]{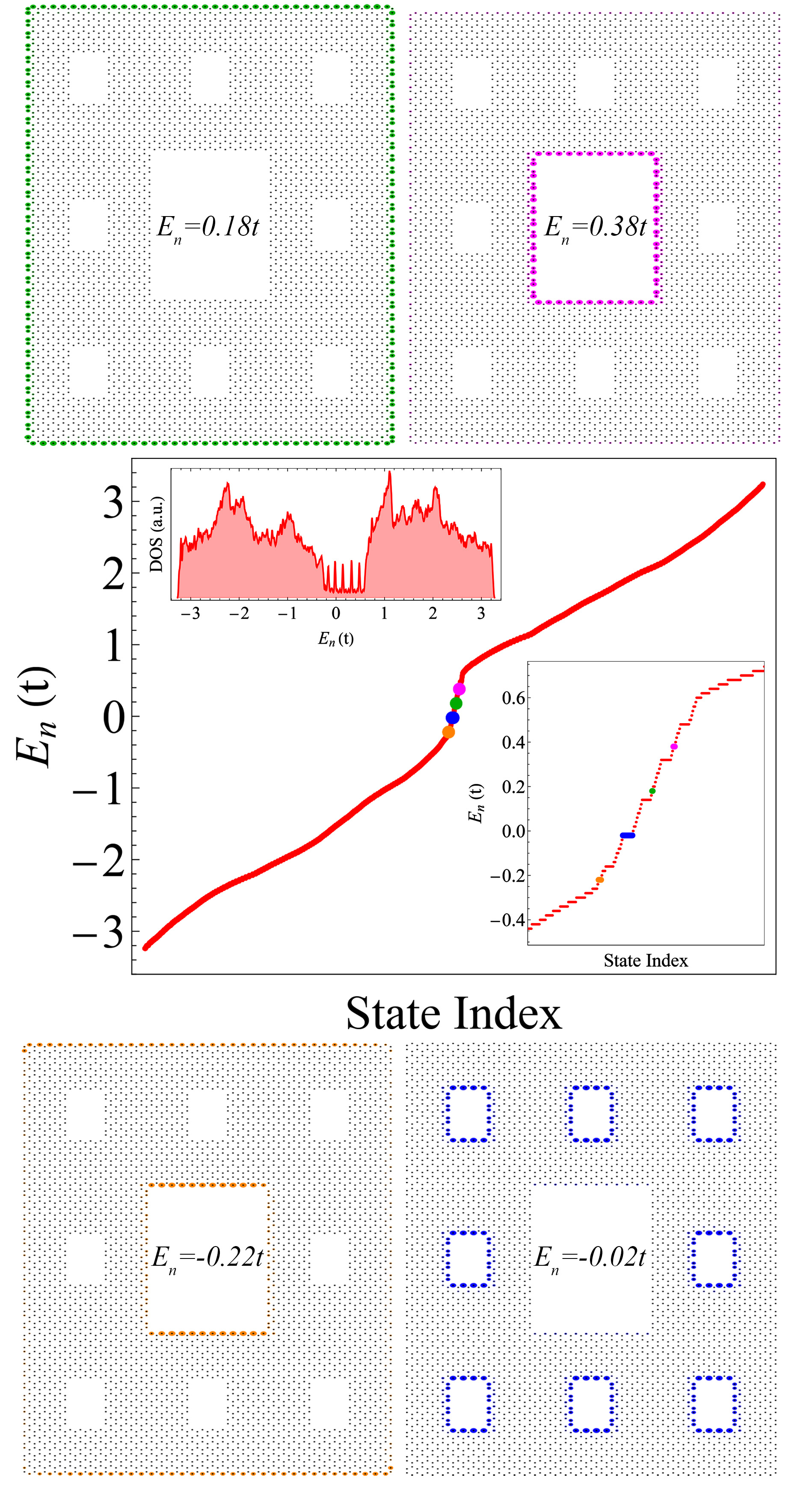}
   \caption{Energy spectra in function of the state index with length of ${\rm dim}(\mathcal{H}_{\rm KMR})=2 \times 4256$. The insets, superior and inferior, for the DOS and zooming the staircase states, respectively. Panels on top and bottom of the energy spectra show the squared modulus of the electronic wave function for in-gap topological states at different energies: $\bar{E}_{n}= -0.22t$, $ -0.02t$, $ 0.18t$, and $ 0.38t$, in orange, blue, green and magenta, respectively. The considered SC G(2) is in the QSHI phase with $\lambda_{\rm KM}=0.25t$, $\lambda_{\rm R}=\sqrt{3}\lambda_{\text{KM}}\approx0.43t$ and $V_{\text{ab}}=0$.}
   \label{figFuncOnda}
\end{figure} 

Figs. \ref{figenergydos}(a) and (b) show the results with model parameters adjusted to match the QSHI phase and the topologically trivial phase, respectively.
The OBC used leads to in-gap topological states when the model's parameters are adjusted to reach a QSHI phase [see Fig. \ref{figenergydos}(a)]. The in-gap states are absent in the topologically trivial situations [Fig. \ref{figenergydos}(b)]. In both the QSHI and trivial phases, the bulk states are not significantly affected by the fractality of the system. The general profile of these energy levels remains nearly unchanged as the fractal generation increases. The respective DOS takes on a noisy appearance with the introduction of fractality. This behavior of bulk energy levels and DOS is consistent with previous studies on SC on a honeycomb mesh, as reported in Refs. \cite{vanVeen2016, Bouzerar2020, Yang2024}. However, the fractal structure drastically alters the in-gap topological states in the QSHI phase, producing a staircase-like profile in this energy spectrum region as the fractal generation increases. This staircase profile shown in the insets of Fig. \ref{figenergydos} (a), translates into sharp peaks in the DOS.

In Fig. \ref{figFuncOnda},  we show the energy spectra of a SC G(2) in the QSHI phase, considering a lower mesh with $N=4256$ sites to avoid computational expenses. It is important to mention that the result captures the physical aspects of wave functions in SCs with denser meshes used in the other numerical results of the present work [Fig. \ref{fig1}]. In fact, in previous work, we showed that the charge distribution converges to a final pattern for a sufficient number of sites \cite{Lage_2025}. We also show in Fig. \ref{figFuncOnda} the wave-function square modulus of topological states at different in-gap energy levels. When an energy level is degenerated due to the staircase profile, we plot the sum of the square moduli of the wave functions in the degenerate subspace and apply a normalization factor.  We highlight two key aspects of the distributions shown in Fig. \ref{figFuncOnda}: (i) The in-gap states closer to the system's neutrality point are localized at the outer edge of SC G(2), as expected from the QSHI picture. The other in-gap states at different energy levels are localized at internal edges formed by the voids introduced in the fractal Sierpinski system. (ii) The distributions obey the $C_2$ symmetry of a SC in the underlying honeycomb mesh, as discussed in Sec. \ref{SecKMRSC}. For the other states outside the in-gap solutions, the LDOS map is distributed regularly over the SC sites, as expected for bulk states.

\subsection{Valence projected spin-spectra}
\begin{figure}[!h]
    \centering
    \includegraphics[width=7cm]{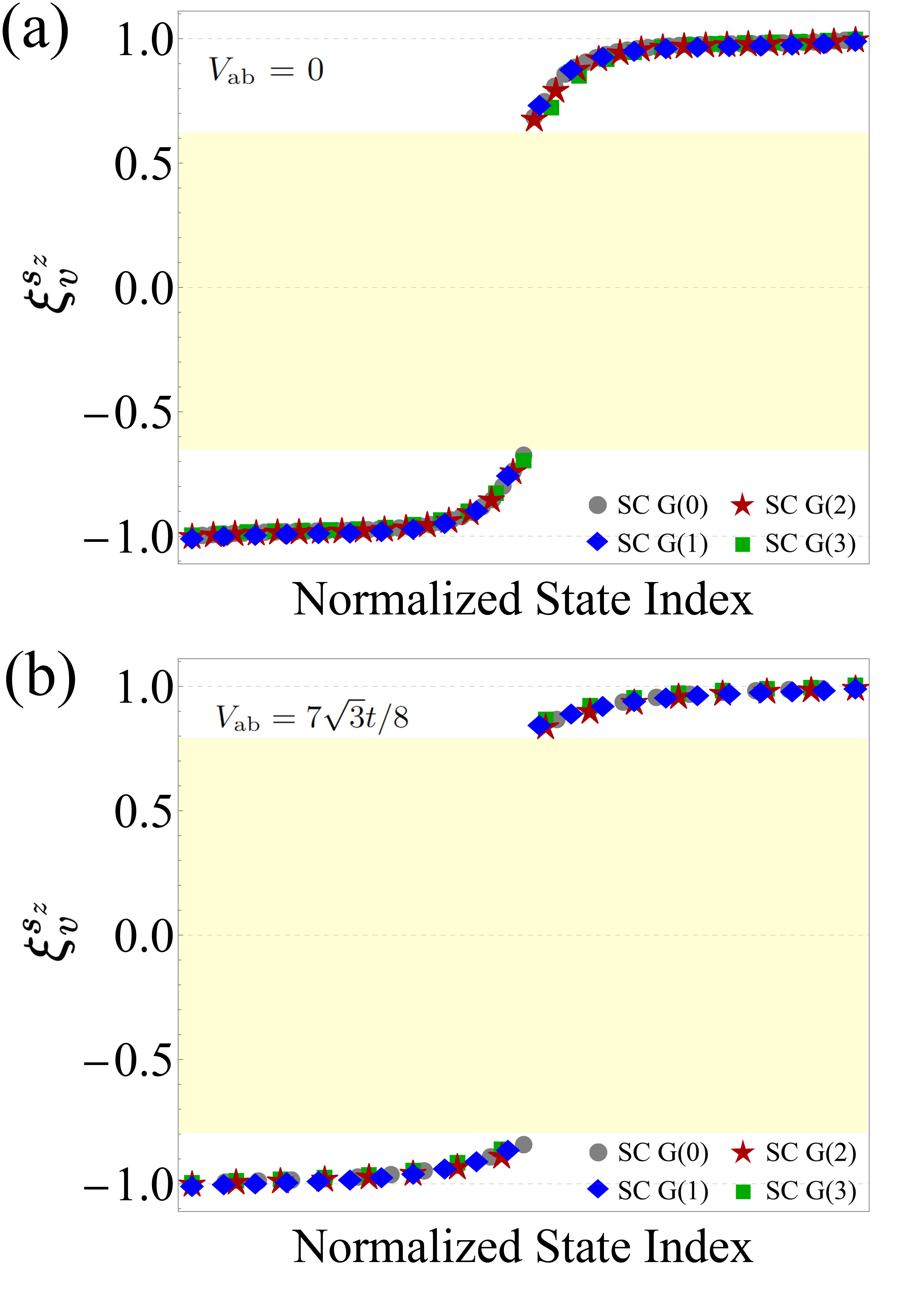}
   \caption{Valence projected spin spectra of the Kane-Mele-Rashba model with $\lambda_{\rm KM}=0.25t$, $\lambda_{\rm R}=\sqrt{3}\lambda_{\rm KM}\approx 0.43 t$ and (a) $V_{\rm ab}=0$ [QSHI phase] and (b) $V_{\rm ab}=7\sqrt{3}t/8\approx 1.51 t$ [trivial phase]. The spectra for different fractal generations is given as a function of the $\text{`normalized state index'}=\text{`state index'}/N_{\rm occ}$. }
   \label{figMs}
\end{figure}
As discussed in Sec. \ref{SecLSCM}, the topological indexation by the LSCM is only possible in the presence of a finite valence-projected spin gap $\Delta^{s_z}$. As previously reported \cite{Prodan2009}, the existence of this gap is not guaranteed by any physical constraint and is associated with the specific details of the model under study . The requirement of a finite $\Delta^{s_z}$ can be seen as a limitation of the approach. Nevertheless, when the valence-projected spin spectra present a finite gap, indexing via the spin Chern number is possible and conveys the same physical information as the $\mathbb{Z}_2$ invariant in time-reversal symmetric systems \cite{Soluyanov2012, Fukui2007, Lange2023}.

In Fig. \ref{figMs}, we present the valence-projected spin spectra $\xi_v^{s_z}$ as a function of the normalized state indexes, for the Kane-Mele-Rashba model in both the QSHI and trivial phases across different fractal generations. The numbers of occupied states at half-filling for the different fractal generations represented in Fig. \ref{fig1} are: $N_{\rm occ}=11988$ for SC G(0), $N_{\rm occ}=10512 $ for SC G(1), $N_{\rm occ}=9456$ for SC G(2), and $N_{\rm occ}=8688 $ for SC G(3). The light-yellow rectangle represents the gap $\Delta^{s_z}$.
One notices that increasing the fractal generation barely alters the valence-projected spin gap $\Delta^{s_z}$ in both the QSHI and trivial phases. This is a remarkable result in this work and means that the LSCM method can be applied to the Kane-Mele-Rashba model at high-fractal generations of the SC. In the next section, we use this approach to study the local topology of the system.

\section{Local topology in SC via LSCM}

The robustness of the valence-projected spin gap $\Delta^{s_z}$ with increasing fractal generation, as presented in Fig. \ref{figMs}, suggests the LSCM method can be applied to locally index the topology of the wave function. This approach was originally introduced in Refs. \cite{Bau2024, Chen2023b, Wieder-Bradlyn-NatCommun} and combines those of Prodan \cite{Prodan2009} and Bianco-Resta \cite{Bianco2011}. It has been shown to accurately capture the topological phase diagram of the Kane-Mele-Rashba model \cite{Bau2024} and its generalizations \cite{tiao}. 

\begin{figure}[!h]
    \centering  \includegraphics[width=9cm]{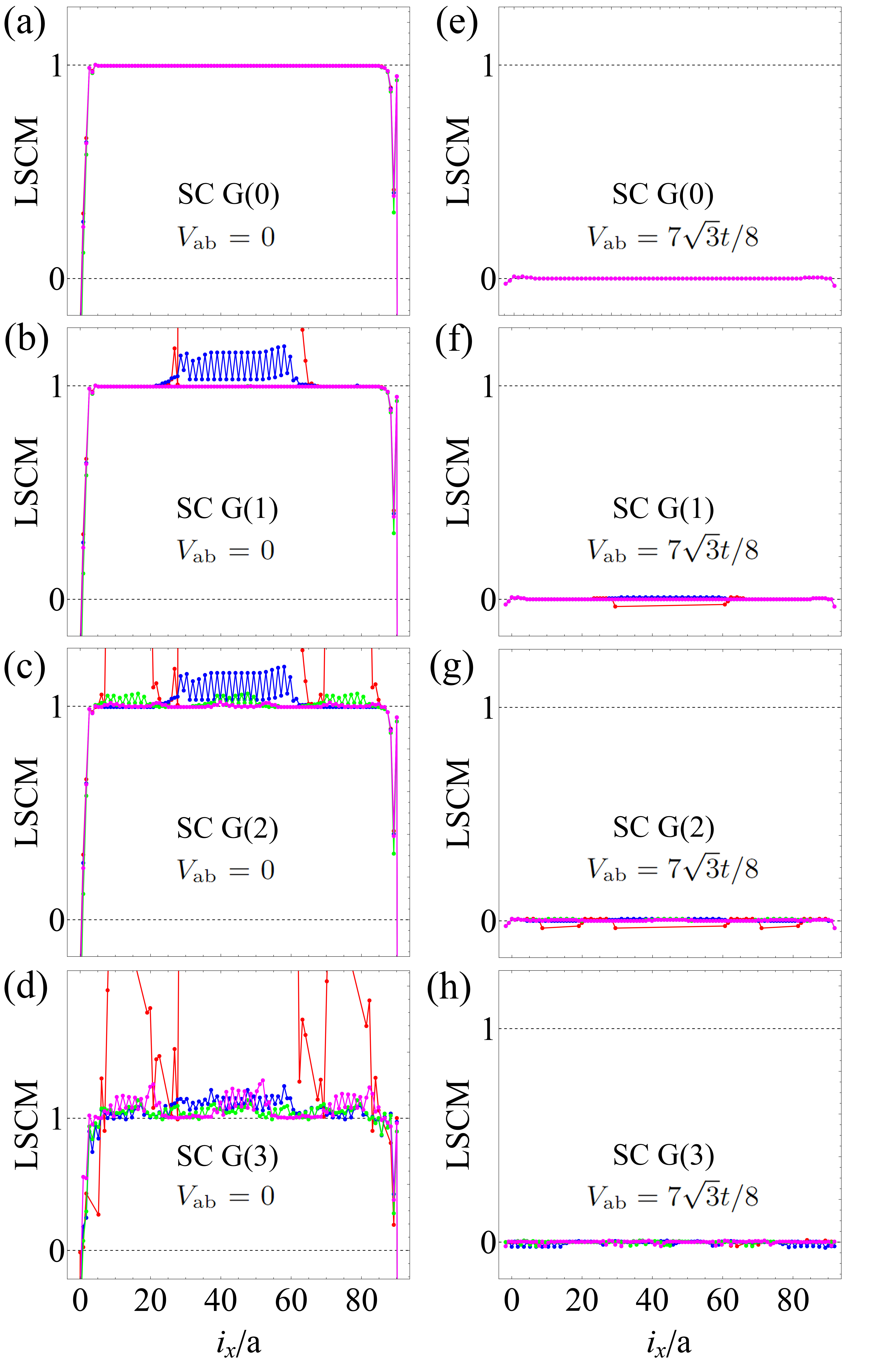}
   \caption{LSCM for SC of G(0-3) (a-d) $V_{ab}=0$ [QSHI phase] and (e-h) $V_{\rm ab} = 7\sqrt{3}t/8$ [trivial phase]. We set $\lambda_{\text{KM}}=0.25t,\lambda_{\text{R}}=\sqrt{3}\lambda_{\text{KM}}$. The colors of the curves follow the color convention of the arrows depicted in Fig. \ref{fig1}: red, blue, magenta, and green curves corresponding to arrows (I), (II), (III), and (IV), respectively. Here, $a$ is the nearest neighbor distance in the honeycomb arrangement.}
   \label{LSCM-lines-QSHI-trivial}
\end{figure}

We show in Fig. \ref{LSCM-lines-QSHI-trivial} the results for the LSCM in the sites along the lines depicted in Fig. \ref{fig1}, marked with different colors, and for the four fractal generations considered. The Hamiltonian parameters in Fig. \ref{LSCM-lines-QSHI-trivial}(a)-(d) were chosen to match the QSHI phase.

For the SC G(0) in the QSHI phase [Fig. \ref{LSCM-lines-QSHI-trivial}(a)] the LSCM along lines (I), (II), (III) and (IV) is quantized for sites in the bulk of the system ${\bf  i}_{\rm bulk}$, $\mathfrak{C}_{s}({\bf i}_{\rm bulk})=1$. Near the edges sites ${\bf i}_{\rm edges}$, the LSCM exhibit well-known anomalies characteristic of local markers \cite{Bianco2011}. Here, we performed our calculations using OBC, but this behavior also occurs in systems with periodic boundary conditions \cite{tiao, WeiChen2023}. Proposals to solve these anomalies involve different enumeration in position operators \cite{Aligia2023, Oliveira2024}. In the case of SC G(1) in the QSHI phase [Fig. \ref{LSCM-lines-QSHI-trivial}(b)], the central void disrupts the quantization of LSCM along lines (I) and (II) in the region $30$ $\scriptstyle{\lesssim}$ ${\bf i}_x/a$ $\scriptstyle{\lesssim}$ $60$. However, the quantization of the marker is preserved along these lines for sites in the regions $5$ $\scriptstyle{\lesssim}$ $ {\bf i}_x/a $ $\scriptstyle{\lesssim}$ $ 30$ and $60$ $\scriptstyle{\lesssim}$ ${\bf i}_x/a$ $\scriptstyle{\lesssim}$ $85$, corresponding to bulk sites of SC G(1).

Lines (III) and (IV) are far from the central void in SC G(1) and are not influenced by its internal edge, thus maintaining the quantization of LSCM in the entire region $5$ $\scriptstyle{\lesssim}$ ${\bf i}_x/a $ $\scriptstyle{\lesssim}$ $85$ enough distant from external edges. As the fractal generation increases by introducing more voids in the system, we observe from Fig. \ref{LSCM-lines-QSHI-trivial}(c) and (d) that regions with anomalies proliferate due to the increasing influence of internal edge regions in sites crossed by lines (I)-(IV). For sites sufficiently far from edges of the system, the LSCM remains quantized. An interesting feature of Figs. \ref{LSCM-lines-QSHI-trivial}(b)–(d) is that the anomalies in the internal edges caused by the voids have the opposite sign compared to those in the external edges.

In Fig. \ref{LSCM-lines-QSHI-trivial}(e)-(h) we report an analogous result but with the Hamiltonian parameter adjusted to match the topologically trivial phase. In this situation, the LSCM takes a value consistent with zero throughout the entire system for different fractal generations. The internal edges introduced by the voids in each fractal generation, merely cause tiny oscillations (of the order $\scriptstyle{\lesssim}$ $10^{-2}$) of the LSCM. This shows that the LSCM also correctly indexes the topologically trivial phase, which is ensured by the existence of a finite spin gap, shown in Fig. \ref{figMs}(b).

\begin{figure}[!h]
    \centering  \includegraphics[width=8.7cm]{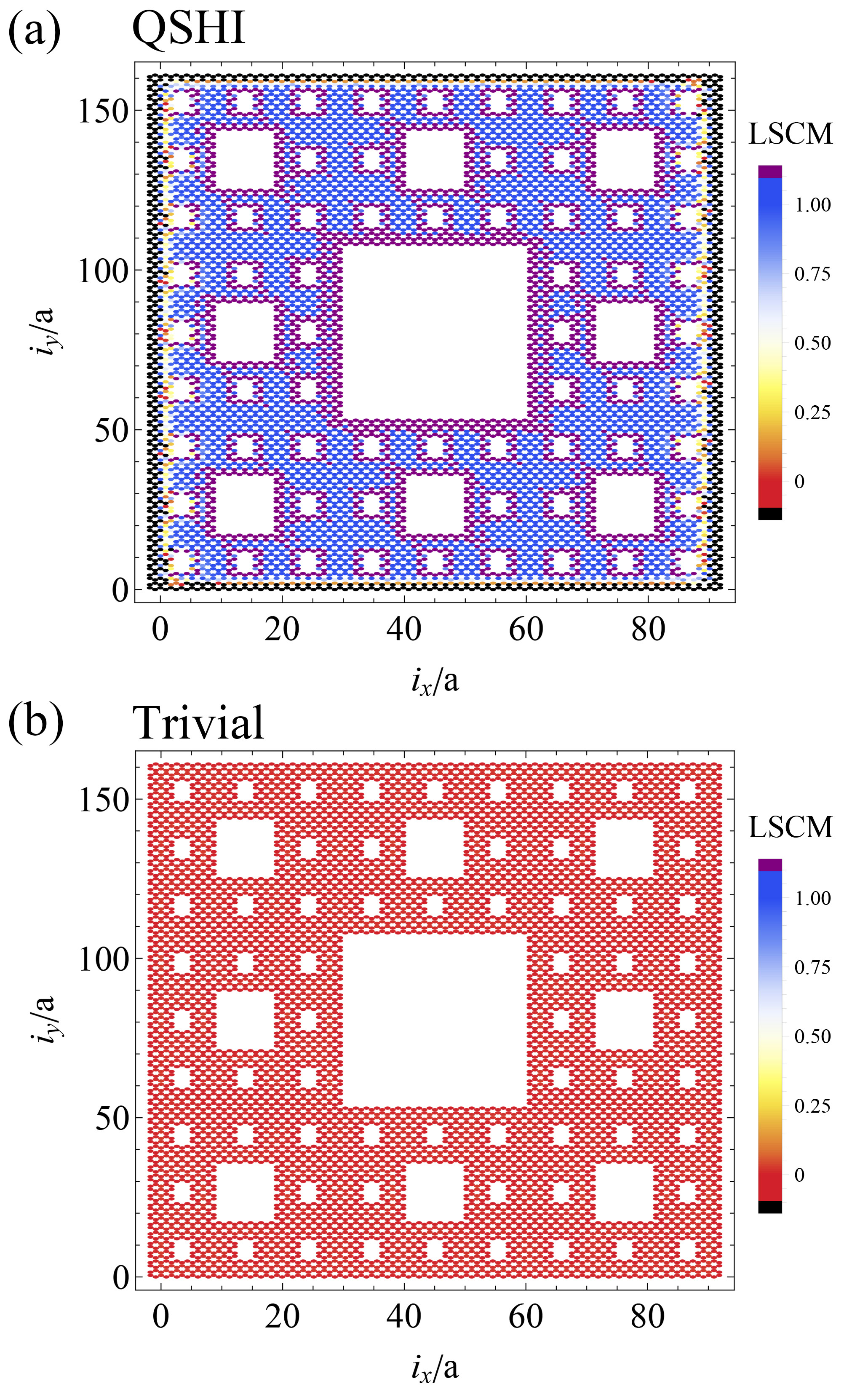} 
   \caption{Site-resolved LSCM for SC G(3). We set $\lambda_{\text{KM}}=0.25t,\lambda_{\text{R}}=\sqrt{3}\lambda_{\text{KM}}$ and sublattice potential  (a) $V_{ab}=0$ [QSHI phase] and (b) $V_{\rm ab} = 7\sqrt{3}t/8$ [trivial phase]. \textcolor{black}{Note that both negative and positive divergences of Fig. \ref{LSCM-lines-QSHI-trivial} (d), marked as black and purple sites in panel (a), respectively, are present in the LSCM spatial map. However, the $\mathfrak{C}_s$ legend color bar is cut at the relevant range, which is between 0 and 1 values.}}
   \label{site-resolved}
\end{figure}

To complement the previous analysis, we present in Fig. \ref{site-resolved} the site-resolved LSCM in the SC G(3) fractal as a color map across the system for both the QSHI [Fig. \ref{site-resolved}(a)] and topologically trivial [Fig. \ref{site-resolved}(b)] phases. One can observe that, despite the presence of many voids in the SC G(3), the LSCM marker remains quantized in a significant portion of the fractal system. For sites located one hexagon away from any edge (internal or external), the quantization of the LSCM $\mathfrak{C}_{s}\approx1$ is achieved in the QSHI phase [Fig. \ref{site-resolved}(a)]. In the trivial phase, one obtains $\mathfrak{C}_{s} \approx 0$ in almost the entire system [Fig. \ref{site-resolved}(b)]. Fig. \ref{site-resolved} shows the ability of this marker to accurately capture the local topological indexation of the system for both phases of the Kane-Mele-Rashba phase diagram in SC fractal geometry. As one approaches the mathematical limit of the SC fractal by increasing the generation further, anomalies caused by internal edges from numerous voids become dominant, compromising the quantization of the marker in QSHI. 

\begin{figure*}[t]
    \centering
    \includegraphics[width=\textwidth]{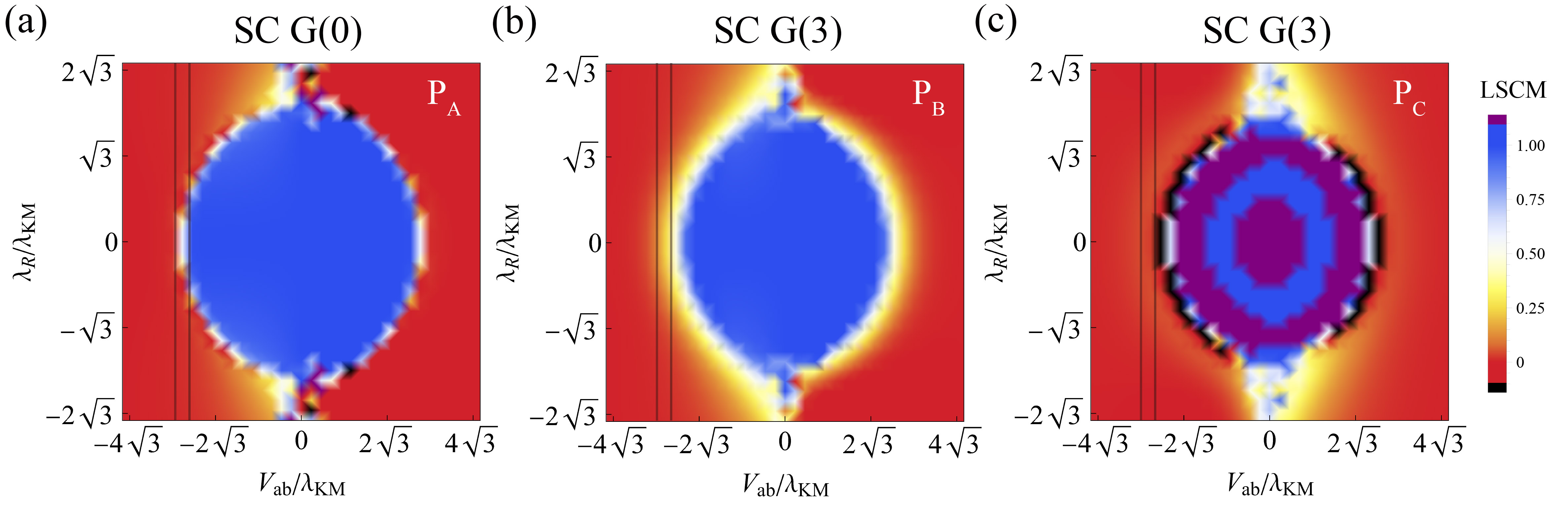}
   \caption{\textcolor{black}{LSCM phase diagrams of the Kane-Mele-Rashba model in the $\lambda_{\rm R}/\lambda_{\rm KM}-V_{\rm ab}/\lambda_{\rm KM}$ space. Here we used $\lambda_{\rm KM}=0.25 t$. In panel (a), we calculate the SC G(0) diagram using the P$_A$ site illustrated in Fig. \ref{fig1}(a) as a reference. In panels (b) and (c), we calculate the SC G(3) diagrams using the P$_B$ and P$_C$ sites illustrated in Fig. \ref{fig1}(d) as references. Vertical black lines serve as guides for the eye, helping to track distortions in the diagrams.}}
   \label{figDiagram}
\end{figure*} 

\textcolor{black}{We now discuss how the Kane-Mele-Rashba phase diagram in parameter space $\lambda_{\rm R}/\lambda_{\rm KM}-V_{\rm ab}/\lambda_{\rm KM}$ is distorted by the fractality. In Fig. \ref{figDiagram}(a), we show the topological phase diagram for SC G(0), using the site P$_{\rm A}$ illustrated in Fig. \ref{fig1}(a). This site lies in the central region of the SC G(0) system and is unaffected by its edges. Therefore, the phase diagram shown in Fig. \ref{figDiagram}(a) closely matches the one obtained from {\bf k}-space calculations (band theory paradigm) \cite{Kane2005b}. In particular, good quantization of LSCM ($\mathfrak{C}_{s}(\text{P}_{\rm A})\approx 1$) is observed in the QSHI region of the phase diagram in Fig. \ref{figDiagram}(a). In the trivial phase, LSCM takes a value close to zero ($\mathfrak{C}_{s}(\text{P}_{\rm A})\approx 0$), as expected. Additionally, a sharp boundary separating the two regions can be seen in Fig. \ref{figDiagram}(a). The phase diagram reported in Fig. \ref{figDiagram}(a) is common to any site chosen in SC G(0) of Fig. \ref{fig1}(a), except those near the external edges.}

\textcolor{black}{If the fractal generation is increased by introducing voids into the system, the LSCM phase diagram can be significantly modified depending on the reference site used in its calculation. We have illustrated the LSCM phase diagram for the SC G(3) fractal in Fig. \ref{figDiagram}(b) and (c), using two different sites, P$_{\rm B}$ and P$_{\rm C}$ (shown in Fig. \ref{fig1} (d)), as references for our calculations. The site P$_{\rm B}$, illustrated in Fig. \ref{fig1}(d), was chosen in a region where the influence of internal edges is minimized. Consequently, the LSCM phase diagram in Fig. \ref{figDiagram}(b) is only slightly distorted compared to the {\bf k}-space diagram. In this case, we observe a reduction in the QSHI phase and a smoothing of the transition between the QSHI and trivial phase regions. Nevertheless, it maintains the expected values for LCSM in the QSHI ($\mathfrak{C}_{s}(\text{P}_{\rm B})\approx 1$) and trivial ($\mathfrak{C}_{s}(\text{P}_{\rm B})\approx 0$) regions, despite the high fractal generation. The site P$_{\rm C}$, also illustrated in Fig. \ref{fig1}(d), was chosen to be strongly influenced by the internal edges of the system. Indeed, the LSCM phase diagram in Fig. \ref{figDiagram}(c) is highly distorted relative to the {\bf k}-space diagram. In particular, the loss of quantization of the LSCM in the QSHI phase indicates a strong influence from anomalies caused by the internal edges of the SC G(3) fractal. Here we also notice a reduction of the QSHI region in the LSCM phase diagram. Note that the different behaviors of the LSCM in SC G(3) at sites P$_{\rm B}$ and P$_{\rm C}$ can be clearly observed in Fig. \ref{site-resolved} (a). Site P$_{\rm C}$ was chosen where the LSCM significantly deviates from the quantized value (purple color in the color map), while site P$_{\rm B}$ was selected where the LSCM exhibits good quantization (blue color in the color map) for the parameters used in Fig. \ref{site-resolved} (a).}

\textcolor{black}{It is noteworthy that at the topological phase transition points (i.e., the boundary between the QSHI and trivial phases), the valence-projected spin gap remains open for the SCs, while the bulk-state energy gap closes. For the topological protection of the marker to be achieved, both gaps must remain open. Therefore, at transition points, one cannot define an LSCM due to the closing of the bulk-state energy gap. As discussed in previous sections, the bulk-state energy gap and the valence-projected spin gap remain open in both the QSHI and trivial phases. Within these phases, the LSCM is well-defined and topologically protected.}

\section{Disordered systems}

\begin{figure*}[h]
    \centering
    \includegraphics[width=0.96\textwidth]{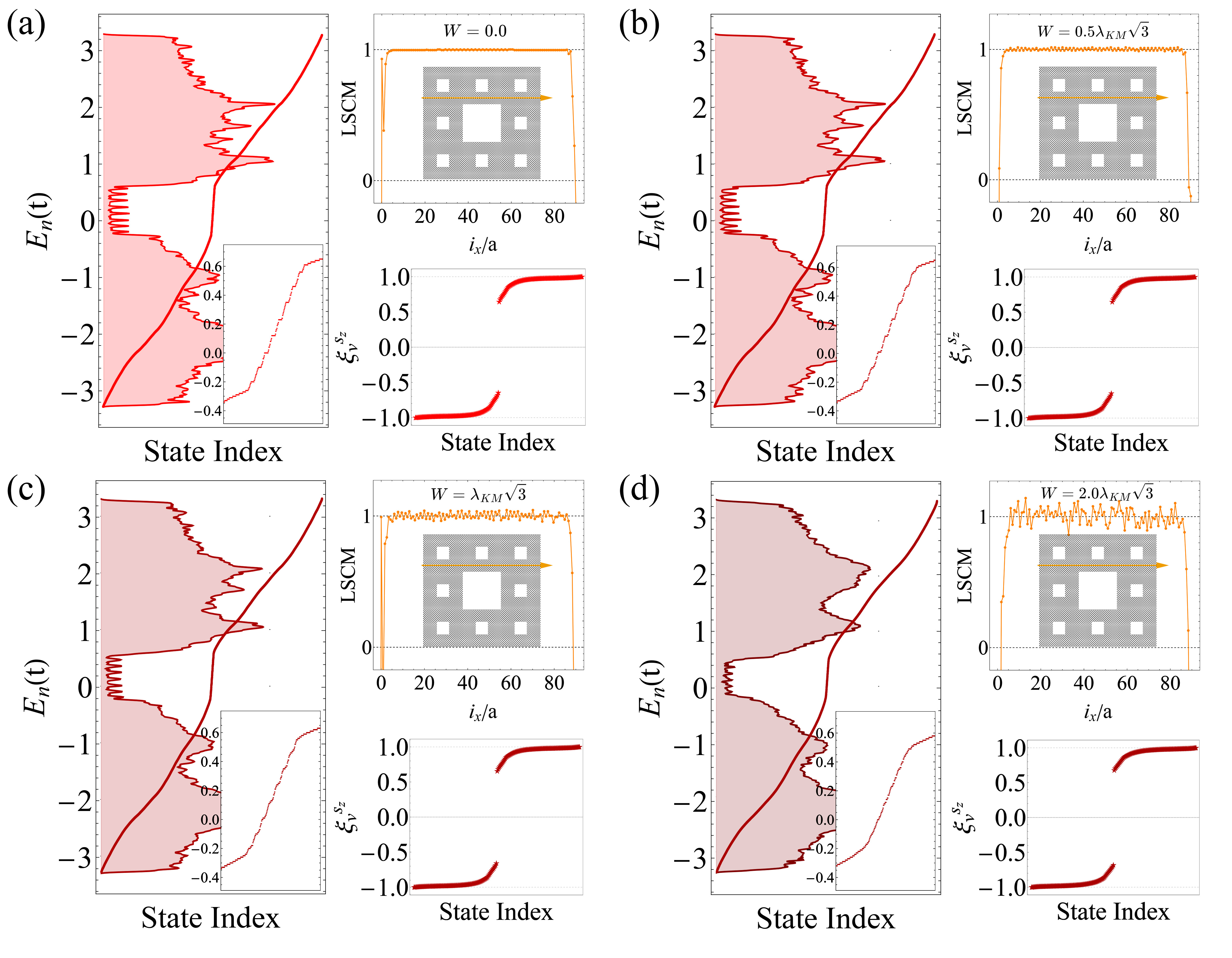}
   \caption{\textcolor{black}{In each panel, we show the following for SC G(2): the LSCM along a line ${\rm i}_x/a$ in the fractal system (golden arrow in the inset); the energy spectra and corresponding DOS; and the valence-projected spin spectra, for different disorder strengths. We also include a zoomed-in view of the region of in-gap energy states to facilitate visualization of the disorder's impact on the staircase profile. Our implementation follows Ref. \cite{Bianco2011} by introducing an on-site disorder energy given by: $\epsilon = W\sum_{s=\uparrow,\downarrow} \sum_{\rm i}\gamma({\rm i})\tau_{\rm i}c^{\dagger}_{{\rm i},s}c_{{\rm i},s} $, where $\gamma({\rm i})$ is a random number uniformly distributed between $0$ and $1$. We set the parameters of the Hamiltonian to drive the system into the QSHI phase: $\lambda_{\rm KM}=0.25 t$, $\lambda_{\rm R}=\sqrt{3}\lambda_{\rm KM}$ and $V_{\rm ab}=0$. The disorder strength values in each panel are (a) $W = 0$, (b) $W = 0.5 \lambda_{\rm KM}\sqrt{3}$, (c) $W = 1.0 \lambda_{\rm KM}\sqrt{3}$, and (d) $W = 2.0 \lambda_{\rm KM}\sqrt{3}$.}}
   \label{figimpurities}
\end{figure*} 

\textcolor{black}{We now discuss the impact of disorder on our previous result. We follow Ref. \cite{Bianco2011}, using a single disorder configuration to qualitatively understand its effects on the local marker and the spectra of the fractal system. It is important to note that this approach provides only a qualitative impression. A more rigorous treatment—including configurational and spatial averaging—would be necessary to fully capture the phenomenology introduced by disorder \cite{Assuncao2024, Oliveira2024, Salib2025}.}

\textcolor{black}{In Fig. \ref{figimpurities}, we present results for the LSCM, the energy spectra, and the valence-projected spin spectra $\xi^{s_z}_v$ of the SC G(2) fractal at various disorder intensities. Here, we focused on a non-trivial topological phase by choosing the parameters of the KMR Hamiltonian to match those of a QSHI. The impact of disorder on the valence-projected spin gap $\Delta^{s_z}$ and on the LSCM quantization (away from the edges) remains relatively weak for disorder intensities up to $W=1.0\lambda_{\rm KM}\sqrt{3}$ [Fig. \ref{figimpurities} (a-c)]. However, for stronger disorder intensities $W=2.0\lambda_{\rm KM}\sqrt{3}$ ($\left| W/L_{\rm b}\right|\approx 0.13$, being $L_{\rm b}$ the band width), the LSCM deviates significantly from its quantized value [Fig. \ref{figimpurities} (d)]. Nevertheless, this deviation may be reduced after taking the configurational average. In contrast, the in-gap energy states and the DOS are strongly influenced by disorder. The peaks in the DOS within the in-gap energy region are progressively smoothed as the disorder strength increases. This occurs due to the fading of the staircase profile in the in-gap energy spectrum. The systematic investigation of how these features vary with configurational and spatial averages in the SC fractal is beyond the scope of this work but is a relevant topic for future studies.}

\section{Final Remarks}

\emph{Comparison to other topological indexes}: Different methods for indexing the topology of fractals have been employed in the literature. Examples include methods based on the Bott index \cite{Lai2024, Li2023, Manna2024} and the real-space version of the global Chern number proposed by Kitaev \cite{Osseweijer2024, Brzezinska2018, Pai2019, Chen2023, Sarangi2021}. As discussed in Ref. \cite{Osseweijer2024}, methods based on the Bott index require the artificial introduction of periodic boundary conditions, which may be inconvenient for fractal structures that typically occur in systems with open boundary conditions. On the other hand, the Chern number method in real space proposed by Kitaev \cite{Kitaev2006} depends on a rather ambiguous and complicated choice of partitioning the system into distinct regions, although it captures the topological aspects of fractals at high generations, when only a small bulk region is available.

\emph{Site-spacing limit}: The study of \textcolor{black}{electronic} fractals in lattice models is inherently constrained by the number of sites used in their construction \cite{Fischer2021}. Starting with a larger number of sites in a SC G(0) structure allows for the introduction of more voids in the system in a self-similar manner before reaching the site-spacing limit. This results in larger fractal generations with a more significant bulk region. As discussed above, a bulk region with at least one hexagon of distance from the internal/external edge of the system is necessary for the marker to present a quantized value in the QSHI phase, for the set of parameters used in previous sections. These limitations mean that computational power is an important issue in the type of study presented in this work.

\textcolor{black}{\emph{Staircase profile in energy spectra}: A remarkable result of this work is the emergence of a staircase profile, induced by fractality, in the in-gap region of the energy spectrum when the model parameters are tuned to the QSHI phase [Fig. \ref{figenergydos}(a)]. Analogous staircase profiles also appear in other contexts related to fractals \cite{Bak1986, Li2023, Wu2022, Jensen1983}. Here we mention the well-known case of confined lattice Fermions subjected to a perpendicular magnetic field. In this situation, the Schr\"{o}dinger equation can be mapped into Harper's equation \cite{Harper1955, Thouless1982}, giving rise to the famous   Hofstadter butterfly spectrum \cite{Hofstadter1976}. The physics of the Hofstadter butterfly arises from the competition between  the lattice spacing and the magnetic length. When the number of magnetic flux quanta per unit cell of the lattice periodic potential is an irrational number, the energy spectrum exhibits self-similar behavior. As discussed in Refs. \cite{denNijs1984, Maska2002}, this system also exhibits a staircase structure and a peaked density of states profile in the energy spectrum.}

\section{Conclusions}
In this work, we present a detailed study of the spectral properties and local topology of the Kane-Mele-Rashba model on a SC fractal constructed over an underlying honeycomb mesh. In the quantum spin Hall insulator phase, the in-gap topological states in the model's energy spectrum exhibit a staircase profile as the fractal generation of the SC increases. This staircase profile in energy spectra translates into the appearance of sharp peaks in the density of states in the in-gap energy region. We also show that the valence-projected spin spectrum exhibits a gap $\Delta^{s_z}$ in both the quantum spin Hall insulator and the trivial phases, which remains robust against an increase in fractal generation. This enables the definition of a local spin Chern marker that captures the spatially resolved topology of the quantum spin Hall insulating phase. We numerically computed the local spin Chern marker in a flake with a large number of sites, showing that this topological marker captures the expected behavior up to the third fractal generation of the Sierpinski carpet. Our work may be useful for the comprehension of topological electronic fractals and expands the scope of local spin Chern marker methods to applications in complex fractal geometry. \textcolor{black}{Additionally, it may provide a solid foundation for future research into spin-dependent quantum transport in fractals \cite{Veen2016, Fischer2021, Fremling2020}, as well as for exploring more intricate fractal geometries \cite{Eek2024}.}

\textcolor{black}{Furthermore, the fractal geometries presented in this work exhibit an intricate spectral distribution. Within the QSHI phase, these fractal states display rich edge configurations with a hierarchical structure. Photonic \cite{Lannebere2019, Lannebere2018, Khanikaev2012}, acoustic \cite{Torrent2012, Khanikaev2015}, and electric circuit \cite{Yang2024, Hofmann2019, Yang2021} analogues of electronic topological models are currently being developed. The results presented here for electronic fractals, particularly the signatures of in-gap edge states in the QSHI phase, could support the design of novel fractal structures in related analogue systems \cite{Biesenthal2022, Zhang2023, He2024}.}

\section*{Acknowledgments}
The authors would like to thank the INCT de Nanomateriais de Carbono for providing support on the computational infrastructure. LLL thanks the CNPq scholarship. SAS-J gratefully acknowledges financial support from the Brazilian Agency CNPq. AL thanks FAPERJ under grant E-26/200.569/2023. The authors thank Rodrigo Arouca for fruitful discussions.

\bibliography{refs}

\end{document}